# MODEL OF ELECTRON CLOUD INSTABILITY IN FERMILAB RECYCLER

S. A. Antipov, The University of Chicago, Chicago, IL 60637, USA
A. Burov, S. Nagaitsev, Fermilab, Batavia, IL 60510, USA

*Abstract*

An electron cloud instability might limit the intensity in the Fermilab Recycler after the PIP-II upgrade. A multi-bunch instability typically develops in the horizontal plane within a hundred turns and, in certain conditions, leads to beam loss. Recent studies have indicated that the instability is caused by an electron cloud, trapped in the Recycler index dipole magnets. We developed an analytical model of an electron cloud driven instability with the electrons trapped in combined function dipoles. The resulting instability growth rate of about 30 revolutions is consistent with experimental observations and qualitatively agrees with the simulation in the PEI code. The model allows an estimation of the instability rate for the future intensity upgrades.

## FAST INSTABILITY

In 2014 a fast transverse instability was observed in the proton beam of the Fermilab Recycler. The instability acts only in the horizontal plane and typically develops in about 20-30 revolutions. It also has the unusual feature of selectively impacting the first batch above the threshold intensity of ~ $4 \cdot 10^{10}$ protons per bunch (Fig. 1). These peculiar features suggest that a possible cause of the instability is electron cloud. Earlier studies by Eldred et. al. [1] indicated the presence of electron cloud in the Recycler and suggested the possibility of its trapping in Recycler beam optics magnets.

The fast instability seems to be severe only during the start-up phase after a shutdown, with significant reduction being observed after beam pipe conditioning [2]. It does not limit the current slip-stacking operation up to 700 kW of beam power, but may pose a challenge for a future PIP-II intensity upgrade [3].

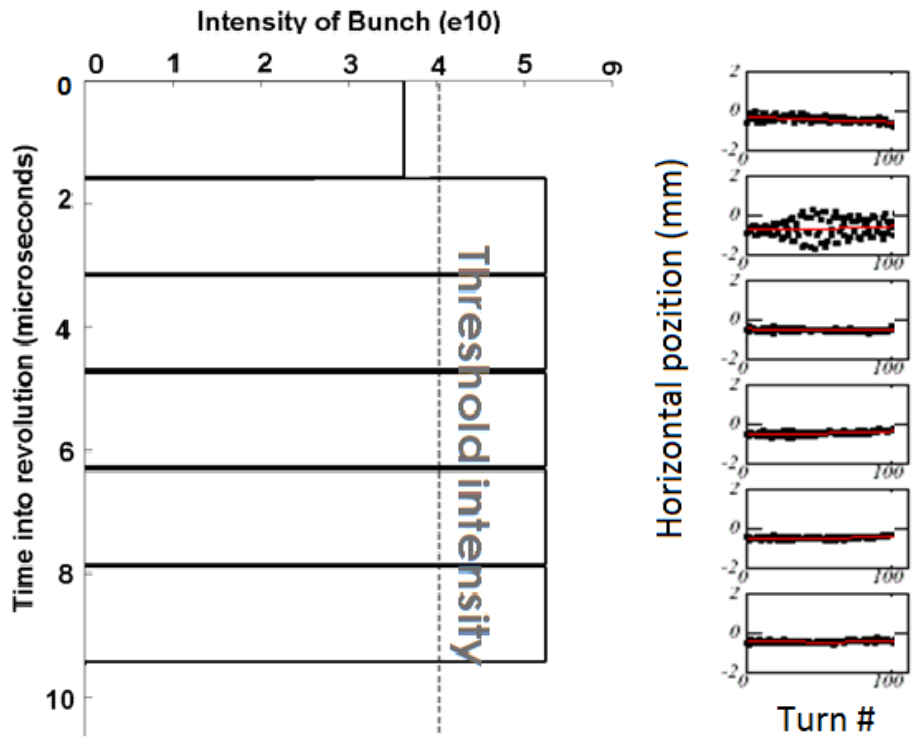

Figure 1: The first batch above the threshold intensity suffers the blow-up after injection into the ring [2].

## ELECTRON CLOUD TRAPPING

The most likely candidates for the source of electron cloud in Recycler are its combined function magnets. They occupy about 50% of the ring's circumference. In a combined function dipole the electrons of the cloud move along the vertical field lines, but the gradient of the field creates a condition for a 'magnetic mirror'– an electron will reflect back at the point of maximum magnetic field if the angle between the electron's velocity and the field lines is greater than:

$$\theta > \theta_{\max} = \cos^{-1}(\sqrt{B_0 / B_{\max}}). \quad (1)$$

Particles with angles $\theta_{\max} < \theta \le \pi / 2$ are trapped by the magnetic field. For Recycler magnets (Table 1), Eq. (1) implies ~$10^{-2}$ of electrons in the cloud are trapped, assuming uniform distribution. A more detailed description of the trapping process is given in [4].

According to numerical studies with the PEI code [4,5], the trapping mechanism allows the electron cloud to gradually build up over multiple turns, reaching a final density orders of magnitude greater than in a pure dipole. The resulting cloud distribution is a stripe along the magnetic field lines, with higher particle density closer to the walls of the vacuum chamber (Fig. 2). The width of the stripe is about the size of the beam. Knowledge of the electron cloud build-up and its distribution allows the construction of a simple model of the electron cloud instability in Recycler.

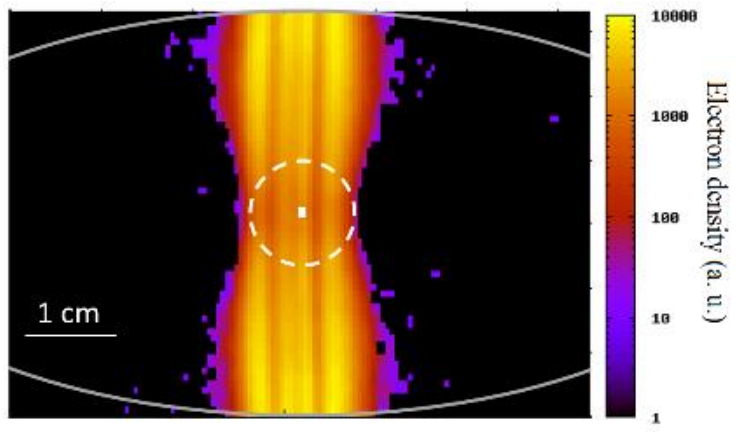

Figure 2: Electron cloud forms a stripe inside the vacuum chamber; the beam center and its 2 rms size are shown in white [4].

## MODEL OF THE INSTABILITY

First, consider a round coasting proton beam travelling in a ring, uniformly filled with electron cloud. Let us denote the position of the beam centroid at an azimuthal angle $\theta$ at time $t$ as $X_p(t, \theta)$. Further, assume that the beam travels at a constant azimuthal velocity around the ring $\omega_0$ and use a smooth focusing approximation with betatron frequency $\omega_\beta$.

For simplicity, we represent the electron cloud by a vertical stripe of uniform charge density, located at a horizontal position $X_e$. Let us further assume that the number of electrons remains, on average, constant in time. Because of the vertical dipole field, the individual electrons of the cloud cannot drift horizontally, but the position of the cloud can change as some regions build up and others are depleted, following the transverse motion of the proton beam. The characteristic time constant of this slow motion is then the time of build-up: $\lambda \sim 1/\tau_{buildup}$.

For small oscillation amplitudes we can assume the electron-proton interaction force to be linear in displacement. Then the coupled collective motion of the beam and the electron cloud is described by the following system of equations:

$$\begin{cases} \left(\frac{\partial}{\partial t}+\omega_0\frac{\partial}{\partial\theta}\right)^2 X_p+\Gamma\left(\frac{\partial}{\partial t}+\omega_0\frac{\partial}{\partial\theta}\right)X_p = \\ \qquad = -\omega_\beta^2 X_p+\omega_p^2 X_e, \\ \frac{\partial}{\partial t}X_e=\lambda(X_p-X_e) \end{cases} \tag{2}$$

where $\Gamma$ is the rate of Landau damping. The coupling frequency $\omega_p$ is

$$\omega_p^2=\frac{\langle N_e\rangle_{circ} r_e c^2}{\pi\sigma^2 R}\cdot\frac{m_e}{\gamma m_p}=\frac{e^2 n_e}{2\varepsilon_0\gamma m_p}, \tag{3}$$

where $r_e$ is the classical radius of electron, $\sigma$ – RMS transverse size of the beam, $R$ – radius of the ring, $n_e$ – electron cloud density, and $\gamma$ – relativistic factor.

The linear damping term $\Gamma$ in the Eq. (2) arises from the spread in betatron frequencies for particles oscillating with different amplitudes. The characteristic rate of the Landau damping can be estimated as

$$\Gamma\sim\omega_\beta\frac{\Delta Q_x}{Q_x}, \tag{4}$$

where $Q_x$ is the horizontal tune and $\Delta Q_x$ is its RMS spread.

Looking for solutions of Eq. (2) in a form $X_{e,p}\propto e^{-i\omega t+in\theta}$ one obtains an equation for the mode frequency $\omega$:

$$-(\omega-\mathrm{n}\,\omega_0)^2-i\Gamma(\omega-\mathrm{n}\,\omega_0)+\omega_\beta{}^2-\omega_p{}^2\frac{i\lambda}{\omega+i\lambda}=0 \tag{5}$$

It can be solved perturbatively, under the assumption that

$$\omega_\beta,\lambda\gg\omega_0,\omega_p,\Gamma. \tag{6}$$

Solving the Eq. (5) in the leading order one gets two modes for each wave number $n$: $\omega_\pm-n\omega_0=\pm\omega_\beta$. Then in first order:

$$\omega_\pm=n\omega_0\pm\omega_\beta+\Delta\omega, \tag{7}$$

where the small complex tune shift $|\Delta\omega|\ll\omega,\lambda,\omega_\beta$ is:

$$\Delta\omega\approx\frac{1}{2}\left[-i\Gamma\mp\frac{\omega_p{}^2}{\omega_\beta}\frac{\lambda(\lambda+i\omega_\pm)}{\lambda^2+\omega_\pm^2}\right] \tag{8}$$

The imaginary tune shift in Eq. (8) consists of two parts with the first being the negative Landau damping term. The motion becomes unstable only if the whole expression is positive. The "+" modes are always stable, while the "-" modes can be unstable for some $n$ if $\mathrm{Im}(\Delta\omega)>0$. The most unstable mode, for which $\mathrm{Im}(\Delta\omega)$ is the greatest, is $\omega_{\max}=\lambda$ and its wave number $n_{\max}$ is

$$n_{\max}=\frac{\omega_\beta+\lambda}{\omega_0}=Q_x+\frac{\lambda}{\omega_0}, \tag{9}$$

and the growth rate of this mode is

$$\gamma_{\max}=\frac{1}{2}\left(\frac{\omega_p{}^2}{2\omega_\beta}-\Gamma\right). \tag{10}$$

In an experiment one will observe the most unstable mode since it quickly suppresses the others due to its rapid exponential growth. The tune shift of this mode is

$$\Delta Q_{\max}\approx\frac{1}{4Q_x}\frac{\omega_p{}^2}{\omega_0{}^2}. \tag{11}$$

This betatron tune shift is what one will observe in an experiment since the most unstable mode quickly suppresses the other modes due to its higher exponential growth rate.

Knowing the complex frequency shift $\Delta\omega$ we can find the impedance of the cloud as (see for example A. Chao [6] Eq. (6.262)):

$$Z=\frac{2\gamma T_0{}^2\omega_\beta}{Nr_0 c}i\Delta\omega, \tag{12}$$

where $N$ is the number of protons in the ring and $r_0$ is the classical proton radius.

Knowing the impedance one can compute the wake functions using formula (2.72) from [6]:

$$W(z)=\frac{-i}{2\pi}\int_{-\infty}^{+\infty}Z(\omega)e^{i\frac{\omega z}{c}}d\omega \tag{13}$$

In the case of a bunched beam, in the rigid bunch approximation, one needs to compute $W(z)$ only at a discreet set of bunch positions $z_k=kc\tau_{rf}$, where $\tau_{rf}$ is the RF period.

Finally, from the impedance of the most unstable mode one can estimate the instability growth rate of a bunched beam as [7]:

$$\gamma_{\max}\approx-\frac{L}{C}\frac{2r_0 N_b\beta_x}{\gamma\tau_{rf}}\mathrm{Re}(Z(\omega_{\max})), \tag{14}$$

where $C$ is the ring circumference and $L$ is the total length of the magnets. For the Recycler $L/C\approx1/2$.

## FAST INSTABILITY IN RECYCLER

Let us apply the model to estimate the parameters of the fast electron cloud instability in Recycler. The input parameters are listed in Table 1. We estimate the density of the electron cloud $n_e\sim10^{12}\,\mathrm{m^{-3}}$ and the rate of its build-up $1/\lambda\sim20\tau_{rf}$. These value were obtained from a numerical simulation and agree with a measurement of the horizontal tune shift [8]. With these parameters the assumption in Eq. (6) holds true.

According to the Eq. (9), the most unstable mode is $n_{\max}\approx30$, and its frequency is about 0.4 MHz. The impedance of this mode, calculated using Eq. (12), is 20 MΩ/m (Fig. 3). Figure 4 depicts the corresponding wake function *W(k)* as a function of bunch number *k*. *W(k)* fits an exponential decay curve

$$W = W_0 \exp(-\Delta z / c\lambda), \Delta z > 0. \qquad (15)$$

The estimate of the mode frequency qualitatively agrees with the simulation in the PEI code and the stripline measurement. PEI simulated the ring, completely filled with 588 bunches of $5\times10^{10}$ p. The resulting frequency is about 0.7 MHz (Fig. 5). In the stripline measurement one batch of 80 bunches of the same charge was injected. The measured frequency was about 0.9 MHz. Both simulated and measured frequencies agree to about a factor of two with each other and the estimate.

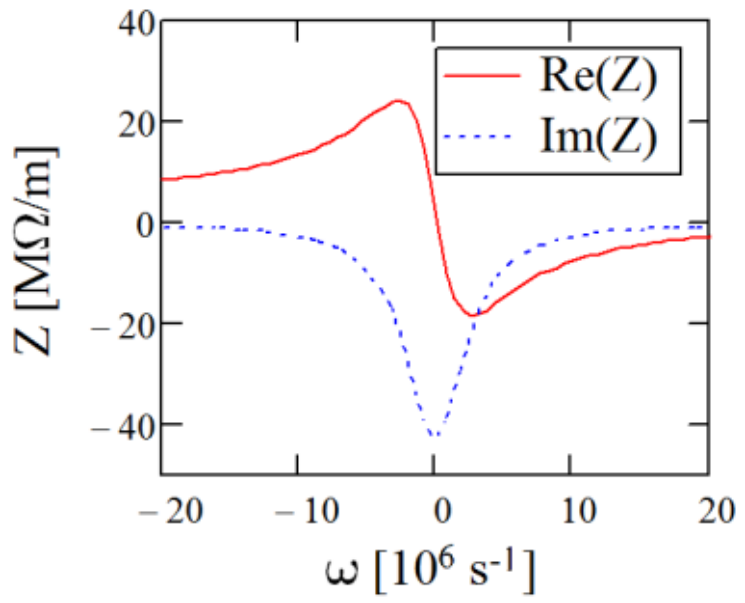

Figure 3: Real and imaginary parts of impedance as a function of a mode angular frequency *ω*.

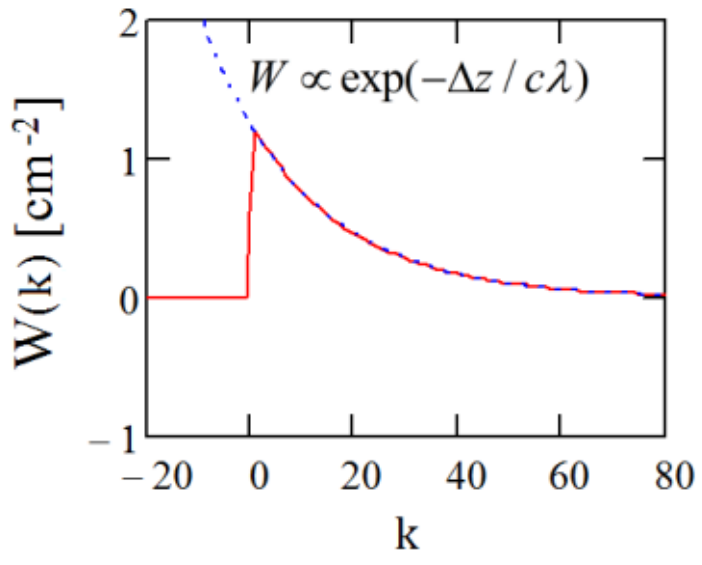

Figure 4: Electron cloud wake falls down exponentially with distance.

Table 1: Parameters of the model

| | |
|---|---|
| Energy, relativistic γ | 8 GeV, 10 |
| Circumference, $\omega_0$ | 3.3 km, $0.57\times10^{6}$ s$^{-1}$ |
| Betatron tune, frequency | 25.45, $14.54\times10^{6}$ s$^{-1}$ |
| Protons per bunch | $5\times10^{10}$ |
| RF harmonic, period | 588, 18.9 ns |
| Electron cloud density | $10^{12}$ m$^{-3}$ |
| e-p coupling frequency $\omega_p$ | $0.23\times10^{6}$ s$^{-1}$ |
| Build-up rate $\lambda$ | $2.65\times10^{6}$ s$^{-1}$ |
| Chromatic tune spread | $2.7\times10^{-3}$ |
| B-field and its gradient | 1.38 kG, 3.4 kG/m |
| Beampipe | Elliptical, 100 x 44 mm |

Using the calculated value of the real part of the impedance we can now estimate the growth rate using Eq. (14). We obtain a growth rate of $\gamma_{\max} = 0.033$ and the characteristic time of the instability $\tau_{\max} = 1/\gamma_{\max} \approx 30$ turns.

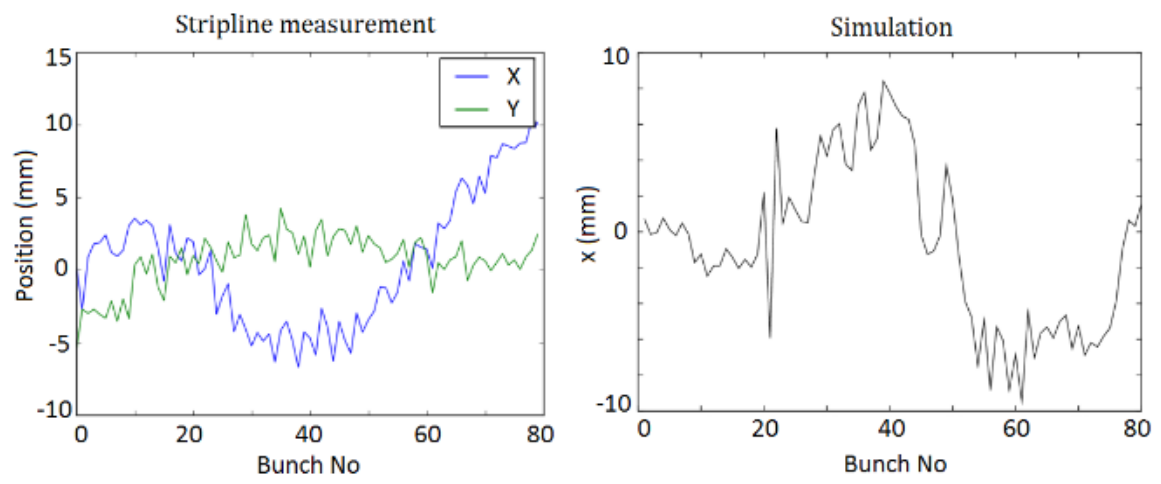

Figure 5: Simulation in PEI and stripline measurements show an instability in the horizontal plane with a period of slightly less than the length of a batch and a frequency < 1 MHz.

## CONCLUSION

A fast transverse instability in the Fermilab Recycler might create a challenge for PIP-II intensities. Understanding its nature is important for making predictions about the machine performance at higher intensities. Earlier studies showed that the instability is likely to be caused by an electron cloud trapped in combined function magnets.

We have constructed a simple analytical model of the electron cloud instability with the cloud trapped in the combined function dipoles. The model allows the estimation of the instability threshold and growth rate for a given electron cloud density.

For the Fermilab Recycler the estimated growth rate is 30 revolutions. This growth rate is consistent with the experimental observations of the instability. The frequency of the most unstable mode of 0.4 MHz is consistent with experimental observations of the fast instability and with numerical simulations in PEI.

The model allows the prediction of the rate of the instability for higher intensities of the proton beam, given an estimate of the electron cloud density, which can be obtained from numerical simulations.

## ACKNOWLEDGMENT

Fermilab is operated by Fermi Research Alliance, LLC under Contract No. DE-AC02-07CH11359 with the United States Department of Energy.

## REFERENCES

[1] J. Eldred, *et al.*, in *Proc. HB'14,* pp. 419-427

[2] J. Eldred, "The High-Power Recycler: Slip-stacking & Electron Cloud", Fermilab, Nov. 2015

[3] S. Holmes, *et al.*, in *Proc. IPAC'16*, pp. 3982-3985

[4] S. A. Antipov, S. Nagaitsev, THPOA49, these proceedings

[5] K. Ohmi, CERN Rep. CERN-2013-002, pp.219-224

[6] A. Chao, *Physics of Collective Beam Instabilities in High Energy Accelerators*, Ney York, NY: Wiley, 1993

[7] Yu. Alexahin, FNAL Beams-doc-4863-v1, Jun. 2015

[8] S. A. Antipov, *et al.*, in *Proc. IPAC'16*, pp. 1728-1730